\documentclass[%
 reprint,
superscriptaddress,
 amsmath,amssymb,
 aps,
 prd,
]{revtex4-2}

\usepackage{physics}
\usepackage{hyperref}
\usepackage{graphicx}
\usepackage{dcolumn}
\usepackage{bm}
\usepackage{capt-of}

\begin{document}

\preprint{APS/123-QED}

\title{Deep Image Reconstruction for Background Subtraction in Heavy-Ion Collisions}

\author{Umar Sohail Qureshi}
 \email{uqureshi@cern.ch}%
\affiliation{Physics Department, Stanford University, Stanford, CA, 94305, USA}
\affiliation{Department of Physics and Astronomy, Vanderbilt University, Nashville, TN, 37235, USA}%
\author{Raghav Kunnawalkam Elayavalli}

\affiliation{Department of Physics and Astronomy, Vanderbilt University, Nashville, TN, 37235, USA}%

\date{\today}

\begin{abstract}
Jet reconstruction in an ultra-relativistic heavy-ion collision suffers from a notoriously large, fluctuating thermal background. Traditional background subtraction methods struggle to remove this soft background while preserving the jet's hard substructure. In this Letter, we present \textsc{DeepSub}, the first machine learning-based approach for full-event background subtraction. \textsc{DeepSub} utilizes a model based on Swin Transformer layers to denoise jet images and disentangle hard jets from the heavy-ion background. \textsc{DeepSub} significantly outperforms existing subtraction techniques by reproducing key jet observables such as jet $p_\mathrm{T}$ and mass, and substructure observables such as girth and the energy correlation function, at the sub-percent to percent level. As such, \textsc{DeepSub} paves the way for precision heavy-ion measurements in hitherto inaccessible kinematic regimes.

\end{abstract}

\maketitle











\setlength{\parindent}{0pt}

\section{Introduction}
\label{introduction}
CERN’s Large Hadron Collider (LHC) and Brookhaven National Laboratory’s Relativistic Heavy-Ion Collider (RHIC) collide ultra-relativistic heavy ions to offer a glimpse at the quark-gluon plasma (QGP), a state of matter where quarks and gluons exist deconfined from hadronic matter.  Arguably, the most powerful probes of the QGP are jets: collimated sprays of particles that result from the fragmentation of high-energy quarks and gluons produced in hard QCD scatterings \cite{Mehtar-Tani:2013pia}. Interactions with the QGP particles modify the jet showering process, thereby leaving the QGP's imprint on the jet itself.

This phenomenon of jet quenching, where highly energetic partons are expected to lose energy while traversing the QGP, has been extensively studied to understand the interaction between jets and the medium \cite{Bjorken:1982tu, dEnterria:2009xfs, Majumder:2010qh, Qin:2015srf}. For instance, studies of jet suppression \cite{PHENIX:2001hpc, STAR:2002ggv, ALICE:2010yje, PHENIX:2003qdj, STAR:2003fka, STAR:2002svs, ALICE:2011gpa, CMS:2012aa, ATLAS:2015qmb, CMS:2016xef}, dijet asymmetry \cite{ATLAS:2010isq, CMS:2011iwn, CMS:2012ulu, CMS:2015hkr}, and splitting functions \cite{CMS:2017qlm, STAR:2021kjt, ALargeIonColliderExperiment:2021mqf}, in addition to modifications in jet shapes \cite{CMS:2013lhm, ALICE:2019whv, ATLAS:2019pid}, fragmentation functions \cite{CMS:2014jjt, ATLAS:2017nre}, and energy-energy correlators \cite{Chen:2024cgx, CMS:2025ydi}, have provided insights into the redistribution of energy and momentum due to QGP interactions. Despite these advances, challenges remain in studying jet quenching due to the steeply falling jet spectrum and even more so, the presence of a large fluctuating background characteristic of heavy-ion environments.

Identifying jet constituents against the background is thus a significant challenge, motivating considerable efforts to develop efficient background subtraction methods. For instance, conventional approaches such as the area \cite{Cacciari_2012, ALICE:2013dpt} and constituent-based subtraction techniques \cite{Berta_2019, Berta:2014eza} estimate and subtract an average (or modulated) background contribution after excluding the leading jets. However, by doing so, they often erase fine-grained jet substructure and consequently produce substantial residuals. Since one cannot identify a single particle as having been produced from either jet or a background, these methodologies often rely on ensemble based approaches that trade background rejection and signal retention. This is particularly important for low-momentum particles in and around jets which are expected to carry signatures of the jet-QGP interaction. These methods, originally developed for proton-proton (pp) collisions, struggle in heavy-ion environments due to much larger backgrounds and non-linear interactions between jets and the QGP. For example, background subtraction methods can be skewed by medium-induced recoil and other noise, distorting both jet momentum and internal structure, an effect that is particularly pronounced for low-$p_\mathrm{T}$ jets. This highlights the need for more sophisticated techniques capable of navigating the complex background while faithfully preserving the detailed substructure of heavy-ion jets.

\begin{figure*}
    \centering
    \includegraphics[width=\linewidth]{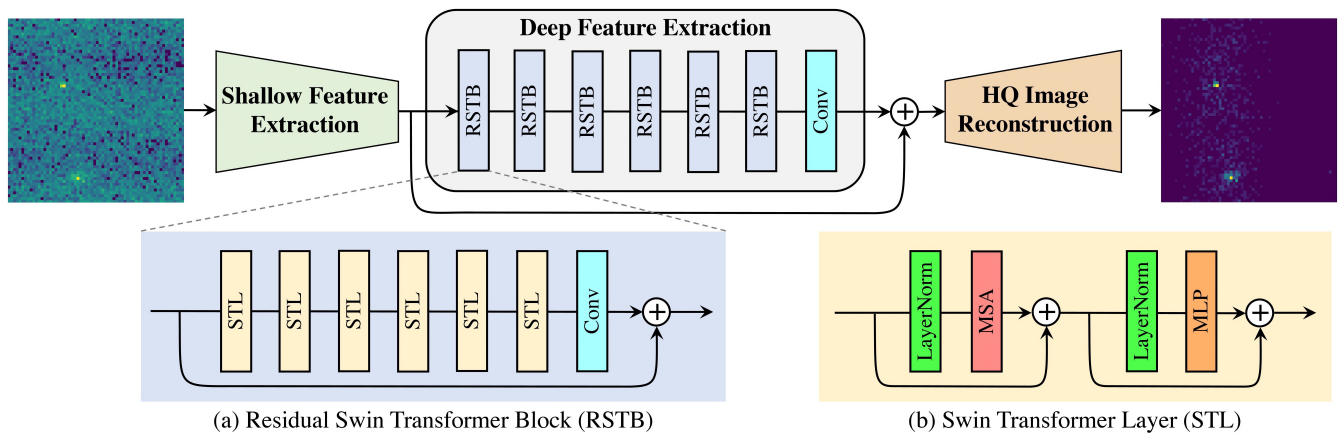}
    \caption{The \textsc{DeepSub} architecture based on the Swin Transformer for image reconstruction.}
    \label{fig:DeepSubarch}
\end{figure*}

The advent of machine learning (ML) has opened new avenues for analyzing heavy-ion collisions \cite{hepmllivingreview}. Recent efforts have leveraged ML on various representations of heavy-ion jets (i.e.~sequences, images, graphs, point clouds, etc.) for quark-gluon discrimination \cite{chien2018probingheavyioncollisions, Du_2022}, tagging quenched jets \cite{Liu:2022hzd, Apolinario:2021olp, Lai:2021ckt, CrispimRomao:2023ssj, Du:2023qst, Qureshi:2024ceh}, predicting jet energy loss \cite{Du_2021}, and locating jet creation vertices \cite{Yang_2023}. However, applications in background subtraction remain limited to reconstructing the transverse momenta \cite{PhysRevC.99.064904, bossi2022rdependenceinclusivejetsuppression, Mengel:2024fcl, li2024jetmomentumreconstructionqgp, mengel2024multiplicitybasedbackgroundsubtraction, du2023overviewjetquenchingmachine} of jets embedded within a heavy-ion background. The aforementioned momentum reconstruction methods are inherently restrictive: they focus on a single one-dimensional observable and thus do not capture the full, high-dimensional structure of jets. In addition, they are often trained on generators that simulate pp collisions and only consider a small dynamic range in the space of jets (e.g.~restricted jet transverse momenta). As such, these methods fail to adapt to the wide range of medium-induced modifications as shown in Refs.~\cite{li2024jetmomentumreconstructionqgp, Stewart:2024mkx}. 

A general-purpose ML-based framework is therefore needed, one capable of learning from the high-dimensional structure jets on an event-by-event basis, encoding non-trivial correlations between signal and background constituents, and robustly disentangling the entirety of the signal instead of extracting a one-dimensional quantity. In this study, we present \textsc{DeepSub}, the first, to our knowledge, ML-based approach for full-event background subtraction. \textsc{DeepSub} takes as input noisy event images, where hard events are embedded in the underlying events, and, outputs the signal image as shown in Fig.~\ref{fig:DeepSubarch}.

The remainder of this Letter is structured as follows. In Sec.~\ref{sampsim}, we describe the simulation procedure for generating hard scattering and the thermal background events. Sec.~\ref{ML} discusses the Swin Transformer-based \textsc{DeepSub} architecture. In Sec.~\ref{res}, we present results on traditionally challenging jet observables and compare them between different background subtraction methods. Finally, we conclude with a short discussion in Sec.~\ref{disc}.

\section{Methodology}

\subsection{Samples and Simulation}
\label{sampsim}
Dijet events are produced considering hard 2-to-2 QCD scatterings with PbPb beams at $\sqrt{s_{\mathrm{NN}}}=5.02$ TeV with $\widehat{p}_\mathrm{T} > 100$ GeV at 0-10\% centrality using \textsc{Jewel} \cite{Zapp:2012ak,Elayavalli_2016, Kunnawalkam_Elayavalli_2017}. No further cutoffs are imposed to ensure that the model is trained on a large dynamic range of jets. To simulate interactions with the QGP medium, we keep recoils on and consider initial temperature $T_i = 590$ MeV and initial quenching time $\tau_i = 0.4$, which provides an adequate description of a variety of jet quenching observables. To remove the energy-momentum contributions from the recoil partons prior to interactions with the hard-scattered partons, the constituent subtraction method \cite{Milhano:2022kzx} is applied.

We simulate background events corresponding to the most‐central (0-10\%) PbPb collisions at $\sqrt{s_\mathrm{NN}} = 5.02$ TeV, in accordance with ALICE data published in Refs.~\cite{ALICE:2016fbt, ALICE:2012nbx, ALICE:2018vuu}. To account for fluctuations in particle multiplicity, we sample the number of particles from a Gaussian distribution with mean $15000$ and standard deviation $122$ (the square root of the mean). The $p_\mathrm{T}$ is sampled from a Boltzmann distribution with mean $\expval{p_\mathrm{T}}$ of 1.2 GeV. The pseudorapidity $\eta$ is sampled uniformly over the interval $\abs{\eta}<3$ and the azimuth $\phi$ from the distribution $f(\phi)=\frac{1}{2\pi}\left[1-v_2 \cos (2\phi)\right]$ with $v_2=0.05$ to account for elliptic flow modulation on $\phi$. Mixed events are subsequently constructed by embedding dijet events into the underlying events.

Event images are then created by discretizing the $\eta{-}\phi$ plane into a $128\times 128$ grid covering $\abs{\eta}<3$ and $0 \le \phi <\pi$. Each final-state particle with coordinates $(\eta_i,\phi_i)$ and transverse momentum $p_{\mathrm{T},i}$ deposits its $p_{\mathrm{T},i}$ into the corresponding pixel, yielding a two-dimensional image of total transverse momentum per bin similar to Refs.~\cite{Cogan_2015, de_Oliveira_2016, Komiske_2017, Lin_2018, Macaluso_2018, Komiske_2017_pumml, Kagan:2020yrm}. For each event, we generate a pair of images: the mixed ``input'' image, $\mathcal{I}_{\mathrm{in}}$, which includes both the underlying event and the dijet, and the ``target'' image, $\mathcal{I}_{\mathrm{target}}$, which contains only the particles from the \textsc{Jewel} dijet.

\subsection{Machine Learning Architecture}
\label{ML}
\textsc{DeepSub} utilizes the Swin Transformer \cite{ liu2021swintransformerhierarchicalvision, liang2021swinirimagerestorationusing} architecture for event image reconstruction as shown in Fig.~\ref{fig:DeepSubarch}. Given an input image $\mathcal{I}_\mathrm{in} \in \mathbb{R}^{H\times W}$, a shallow feature extraction layer $h_{\mathrm{SF}}$ consisting of a $3\times 3$ convolutional layer, extracts shallow features $F_0 \in \mathbb{R}^{H\times W \times C}$ as $F_0 = h_{\mathrm{SF}}(\mathcal{I}_\mathrm{in})$, where $C$ is the latent feature dimension. Deep feature representations $F_{\mathrm{DF}}\in \mathbb{R}^{H\times W \times C}$ are then extracted from $F_0$ as $F_{\mathrm{DF}} = h_{\mathrm{DF}}(F_0)$ with $h_{\mathrm{DF}}$ denoting the deep feature extraction module which consists of $K$ Residual Swin Transformer Blocks (RSTBs) followed by a $3\times 3$ convolution. In particular, intermediate features $F_1, F_2, \dots, F_K$ and $F_{\mathrm{DF}}$ are constructed sequentially as $F_i = h_{\mathrm{RSTB}}^i(F_{i-1)}$ for $i \in \{1, 2, \dots, K\}$ and $F_\mathrm{DF}=h_\mathrm{conv}(F_K)$ where $h_{\mathrm{RSTB}}^i$ and $h_\mathrm{conv}$ denote the $i^\mathrm{th}$ RSTB and final convolution layer respectively.

As the name suggests, the RSTB is a residual block with Swin Transformer Layers (STLs) followed by a convolution. Given the input feature $F^i_0$ of the $i^\mathrm{th}$ RSTB, the features $F^i_1, F^i_2, \dots, F^i_L$ by $L$ STLs are computed as $F_j^i = h_{\mathrm{STL}_j^i}(F_{j-1}^i) $ for $j\in\{1,2, \dots, L\}$ where $h_{\mathrm{STL}_j^i}$ is the $j^\mathrm{th}$ STL in the $i^\mathrm{th}$ RSTB. Finally, the output is generated by applying a convolution before the residual connection: $F^i_{\mathrm{out}} = h_{\mathrm{conv}_i}(F^i_L)+F^i_0$ where $h_{\mathrm{conv}_i}$ is the final convolutional layer in the $i^\mathrm{th}$ RSTB. Finally, the shallow and deep features are convolved with a single convolutional layer to produce the reconstructed output image $\mathcal{I}_{\mathrm{reco}}$ as $\mathcal{I}_\mathrm{reco}=h_{\mathrm{conv}_{\mathrm{IR}}}(F_0+F_{\mathrm{DF}})$ where $h_{\mathrm{conv}_{\mathrm{IR}}}$ is the convolution layer in the HQ Image Reconstruction module.

The STL extends the original multi‐head self‐attention (MSA) mechanism of the standard Transformer \cite{vaswani2023attentionneed} by restricting attention to local windows and alternating these windows in a shifted configuration to enable cross‐window interactions. Given an image $\mathcal{I} \in \mathbb{R}^{H\times W\times C}$, it is first partitioned into $\tfrac{HW}{M^2}$ non‐overlapping windows of size $M\times M$, each reshaped into a feature matrix $X\in\mathbb R^{M^2\times C}$. Within each window, projections $P_Q,P_K,P_V\in\mathbb R^{C\times d}$ produce queries $Q=XP_Q$, keys $K=XP_K$, and values $V=XP_V$ in $\mathbb R^{M^2\times d}$. Local self‐attention is subsequently computed as $\mathrm{SoftMax}\left(\frac{QK^\top}{\sqrt d}+B\right)V$ where $B$ denotes learnable relative positional biases. Multi‐head self‐attention concatenates $h$ such attention outputs computed in parallel. A subsequent two‐layer MLP with GELU activation is further applied to the features. Both the attention and MLP blocks employ pre‐LayerNorm and residual connections, thus giving $\widetilde X=\mathrm{MSA}(\mathrm{LN}(X))+X$ and then $X'=\mathrm{MLP}(\mathrm{LN}(\widetilde{ X}))+\widetilde{ X}$. In contrast to the Vision Transformer (ViT) \cite{dosovitskiy2021imageworth16x16words}, which applies global self‐attention over all tokens with $\mathcal O\big(H^2W^2\big)$ complexity, the Swin Transformer’s hierarchical, window‐based design achieves linear $\mathcal O\big(HWM^2\big)$ complexity while still capturing long‐range dependencies via its shifted window scheme as well as bringing in strong inductive biases for local translational equivariance.

\begin{figure*}
    \centering
    \begin{minipage}{0.495\textwidth}
        \includegraphics[width=\linewidth]{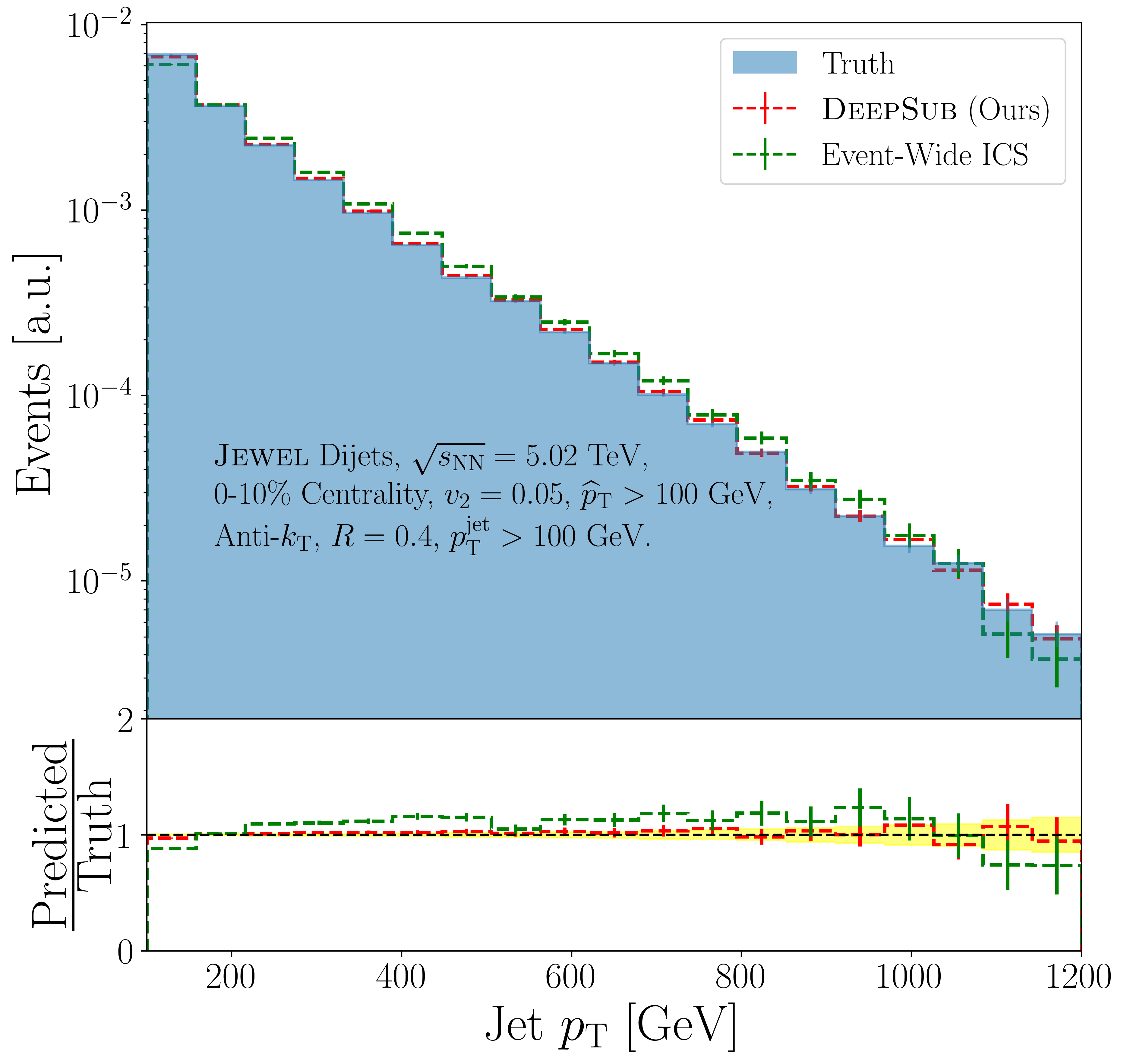}
    \end{minipage}
    \begin{minipage}{0.495\textwidth}
        \includegraphics[width=\linewidth]{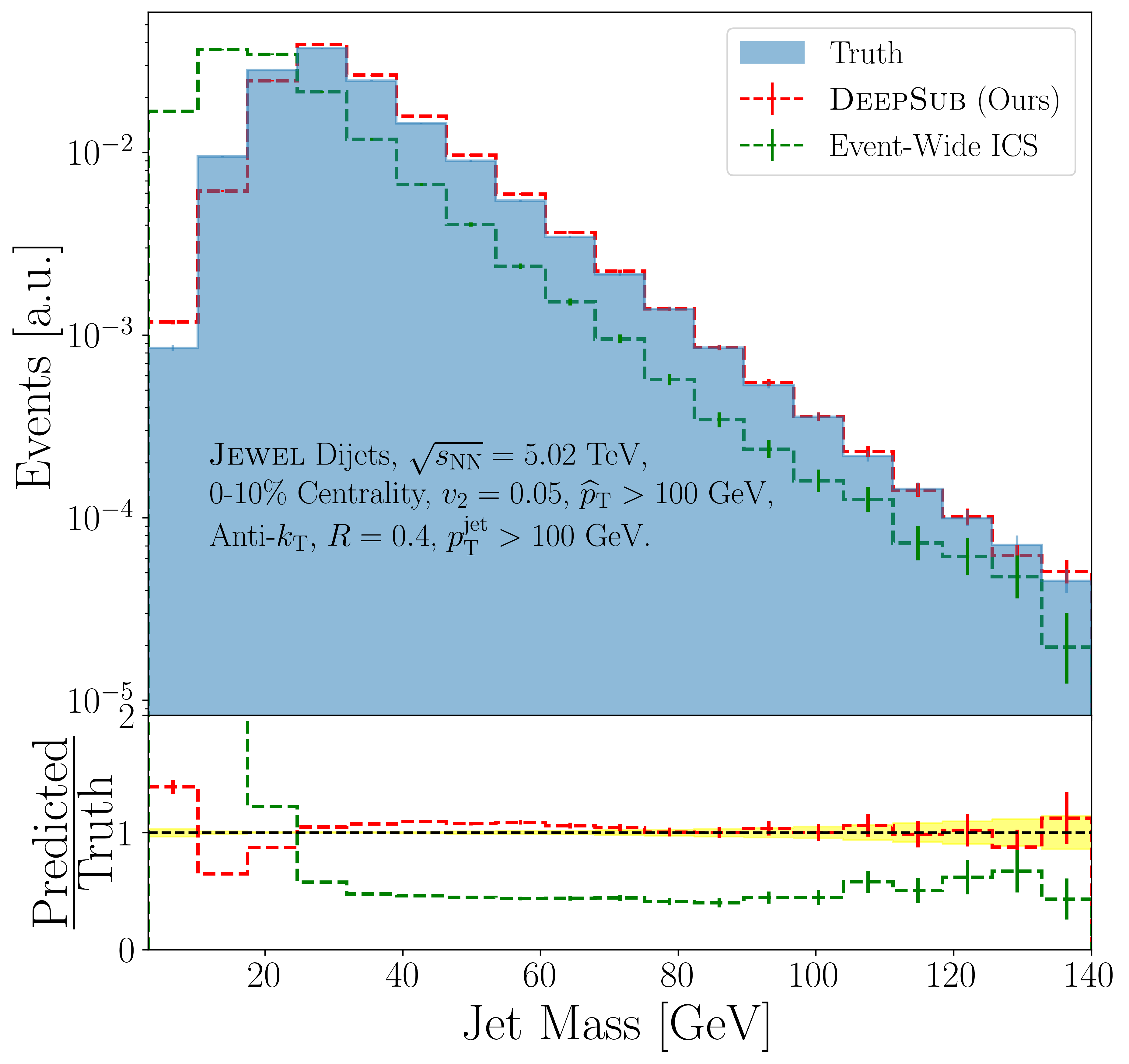}
    \end{minipage}
    \begin{minipage}{0.495\textwidth}
        \includegraphics[width=\linewidth]{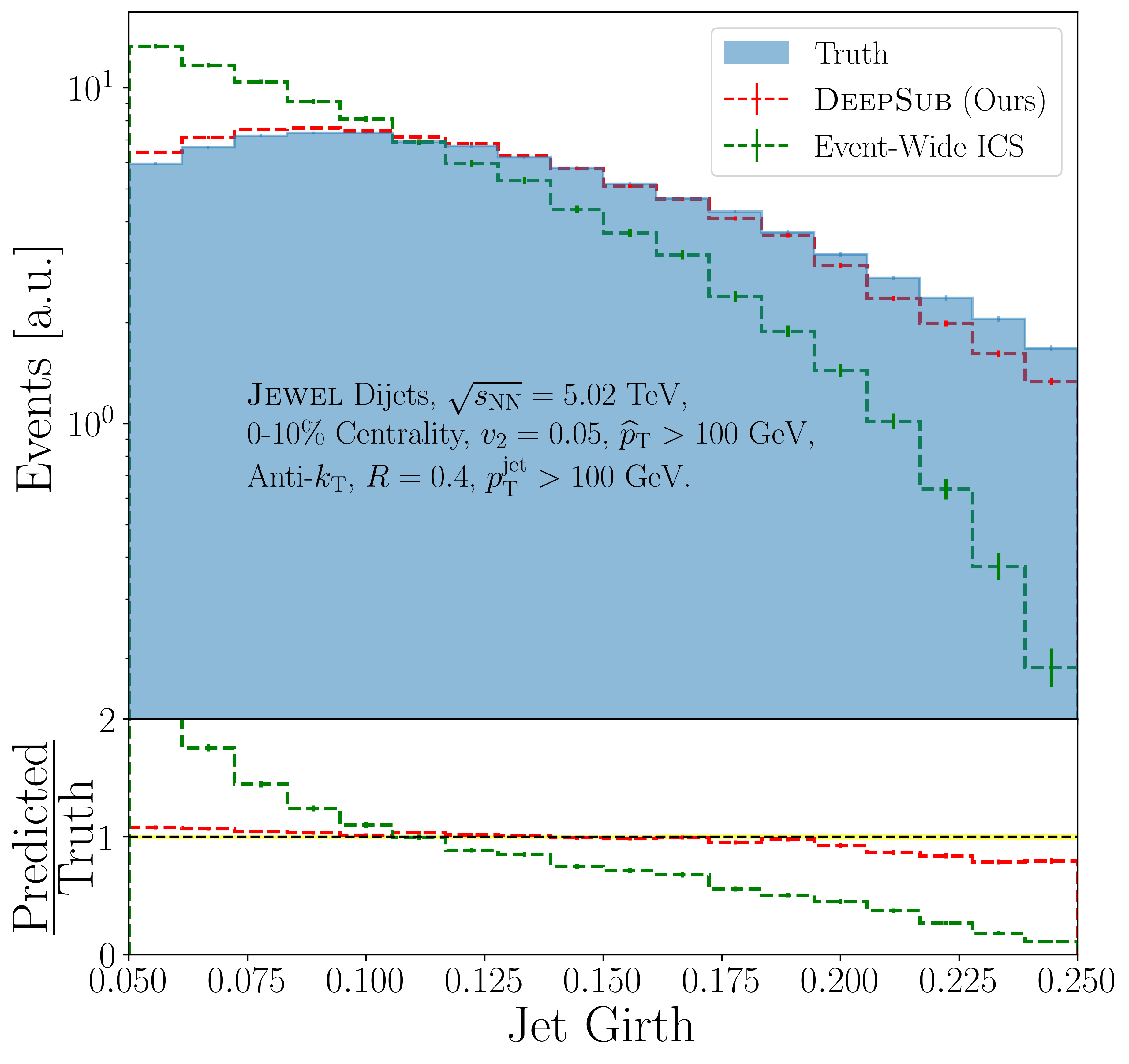}
    \end{minipage}
    \begin{minipage}{0.495\textwidth}
        \includegraphics[width=\linewidth]{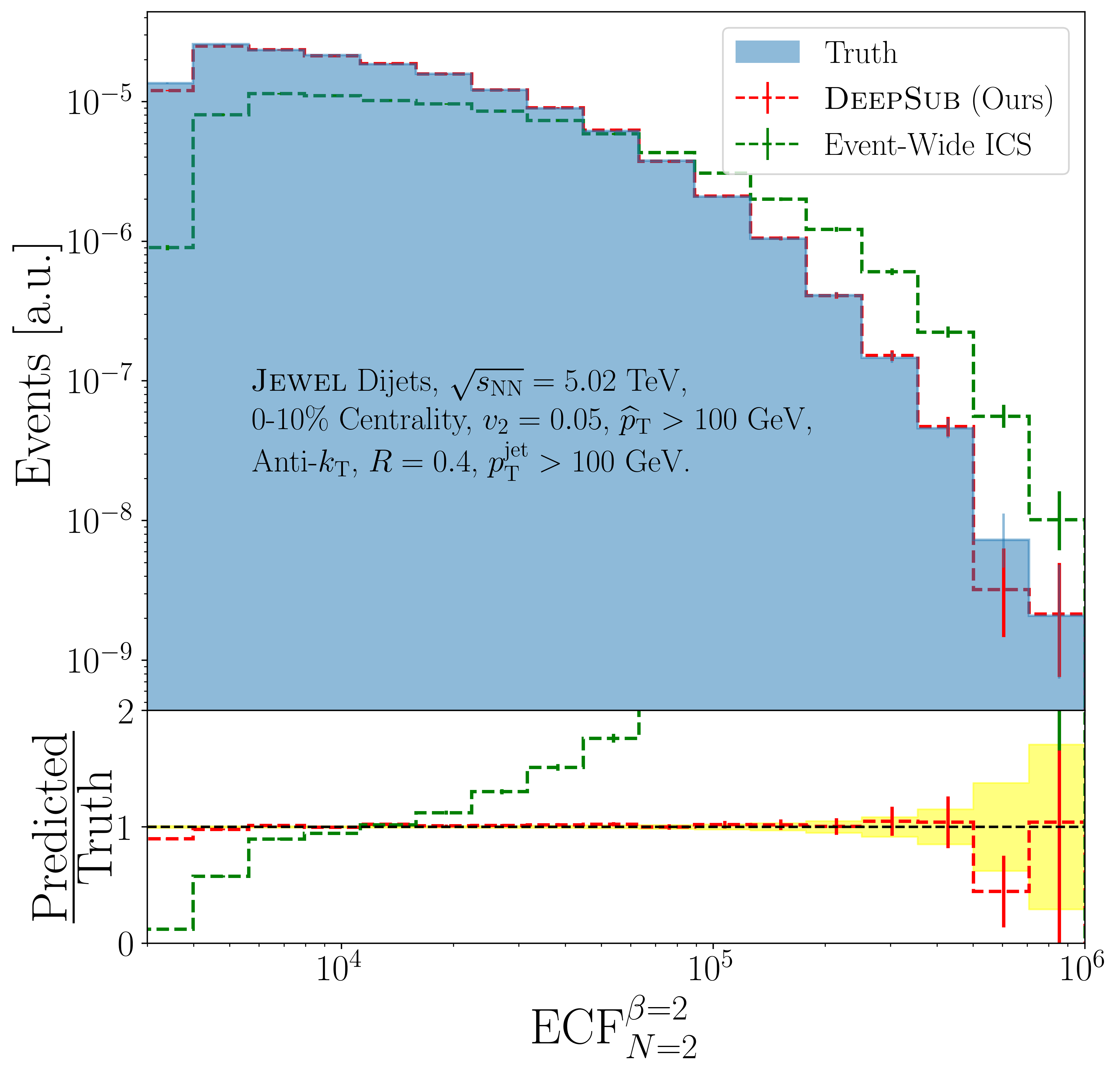}
    \end{minipage}
    \caption{Distributions of jet $p_\mathrm{T}$ (top-left), mass (top-right), girth (bottom-left), and the energy correlation function (bottom-right) for \textsc{DeepSub} (red-dotted), event-wide ICS (green-dashed), and the ground truth (blue-filled) on \textsc{Jewel} dijet events. The error bars represent the statistical uncertainty in each bin. The bottom panels in each plot show the ratio of the reconstructed value of each observable to their truth values and the yellow-shaded region indicates the statistical uncertainty in the truth.}
    \label{fig:resJewel}
\end{figure*}

\section{Results}
\label{res}
For this study, the number of RSTBs, STLs, and attention heads is kept to 6 each. The window size and latent feature dimension are kept to 8 and 180, respectively. The model in this formulation has 11.5M parameters and uses roughly 5 GFLOPs on the forward pass. The canonical \textsc{PyTorch} \cite{paszke2019pytorchimperativestylehighperformance} deep learning framework was employed for implementing, configuring, training, and evaluating the model on 4 $\times$ Nvidia RTX A100 GPUs. The \textsc{Adam} \cite{kingma2017adammethodstochasticoptimization} optimizer with a learning rate of $7\times 10^{-5}$ is used in conjunction with the mean-squared error (MSE) loss function between $\mathcal{I}_{\mathrm{reco}}$ and $\mathcal{I}_{\mathrm{target}}$. A train-validation-test split consisting of 360k, 40k, and 80k events, respectively, is used with a batch size of $128$, and the model is trained for 15 epochs until convergence. The best model based on the validation set MSE loss is used for all subsequent testing.

The output space of events generated by \textsc{DeepSub} is too high-dimensional to be examined holistically. Therefore, to evaluate its performance, we compute a few observables and compare their truth values to the reconstructed values. We compare \textsc{DeepSub} to the powerful and widely used event-wide iterative constituent subtraction (ICS) method \cite{Berta:2014eza} with two iterations using distance parameters $\Delta R_{\text{max}} = \{0.2, 0.1\}$ between signal and background particles, and a transverse momentum weight $\alpha = 1$. Jets are clustered after background subtraction with the anti-$k_\mathrm{T}$ algorithm \cite{Cacciari:2008gp} requiring $R = 0.4$, $p^{\mathrm{cons}}_\mathrm{T}>0.3$ GeV, $p_\mathrm{T}^{\mathrm{jet}} > 100$ GeV, and $\abs{\eta} < 2.5$ using \textsc{FastJet} \cite{Cacciari:2011ma, Cacciari:2005hq, Roy_2023}. All particles are given the charged pion $\pi^\pm$ mass of $139.5$ MeV. Our comparisons focus on the following four jet observables: (i) jet $p_\mathrm{T}$ i.e.~the transverse momentum of the jet; (ii) jet mass i.e.~the Lorentz-invariant mass of the four-momentum sum of all constituent particles; (iii) the jet girth $g$ defined as:
\begin{equation}
    g = \frac{1}{p_\mathrm{T}^{\mathrm{jet}}}\sum_{i\in \mathrm{jet}} p_{\mathrm{T}}^i \Delta R(i, \mathrm{jet}),
\end{equation}
Where $p_{\mathrm{T}}^i$ is the transverse momentum of constituent $i$ and $\Delta R(i, \mathrm{jet})$ is the distance to the jet axis; and (iv) the energy correlation function (ECF) \cite{Larkoski_2013} with $N=2$ and $\beta=2$ defined as:
\begin{equation}
    \mathrm{ECF}_{N=2}^{\beta=2} = \sum_{i\in \mathrm{jet}}\sum_{j<i} p_{\mathrm{T}}^i p_{\mathrm{T}}^j \Delta R(i, j)^2.
\end{equation}

The top panels of Fig.~\ref{fig:resJewel} illustrate the distributions of these observables after applying the two background subtraction methods. A more useful comparison of performance is shown in the bottom panels of each plot i.e. the ratio of the reconstructed value of each observable to their truth values. From a quick glance at the plots, it is clear that \textsc{DeepSub} more closely matches, by a significant margin, the true distributions of more sensitive substructure observables like mass, girth, and the energy correlation functions as compared to the event-wide ICS method. 

The top-left plot shows jet $p_\mathrm{T}$, which falls by nearly four orders of magnitude between 100 and 1200 GeV in a characteristic power‐law. Overlaid are the background‐subtracted spectra: \textsc{DeepSub} (red-dashed) and the event-wide ICS method (green-dashed). Both closely follow the true distribution, but with small deviations. In the 100-200 GeV region, ICS overshoots the truth by $\approx5\%$, while \textsc{DeepSub} stays essentially on top of it. From 200 to 800 GeV, ICS continues to exceed the truth by up to $\approx 10\%$, whereas \textsc{DeepSub} remains within $ 1\%$ of unity. Above $\approx 1$ TeV, both methods fluctuate about the true yield, ICS by as much as $\approx 20\%$ and \textsc{DeepSub} by $\approx 5\%$, though much of these can be attributed to statistical uncertainties. Overall, \textsc{DeepSub} delivers a more faithful reconstruction of the hard‐jet spectrum with smaller residual deviations compared to the traditional event-wide ICS approach.

The top-right panel shows the reconstructed jet mass spectrum, which falls by roughly three orders of magnitude. Both methods struggle at the very lowest mass bins ($\approx$ 3-20) GeV, here ICS overshoots the truth by as much an order of magnitude, while \textsc{DeepSub} overpredicts by $\approx 40\%$. In the intermediate mass region (20–60 GeV), ICS underestimates the yield by a factor of $\approx$ 2-3, whereas \textsc{DeepSub} remains within 1{-}5\% of the true spectrum. At high masses ($>80$ GeV), ICS still runs low by a factor of 2 and \textsc{DeepSub} by $\approx$ 1-5\% except in the very last bins where statistical uncertainties dominate. Since jet mass is largely governed by large-angle structure, this behavior is indicative of over-subtraction and destruction of genuine large-angle soft radiation within the jet by the ICS method, a notion that is supported by Refs.~\cite{Berta_2019, Berta_2014}. In contrast, \textsc{DeepSub} is able to preserve this softer wide-angle information and delivers a markedly more accurate reconstruction of the jet mass spectrum than ICS.

\begin{figure*}
    \centering
    \begin{minipage}{0.495\textwidth}
        \includegraphics[width=\linewidth]{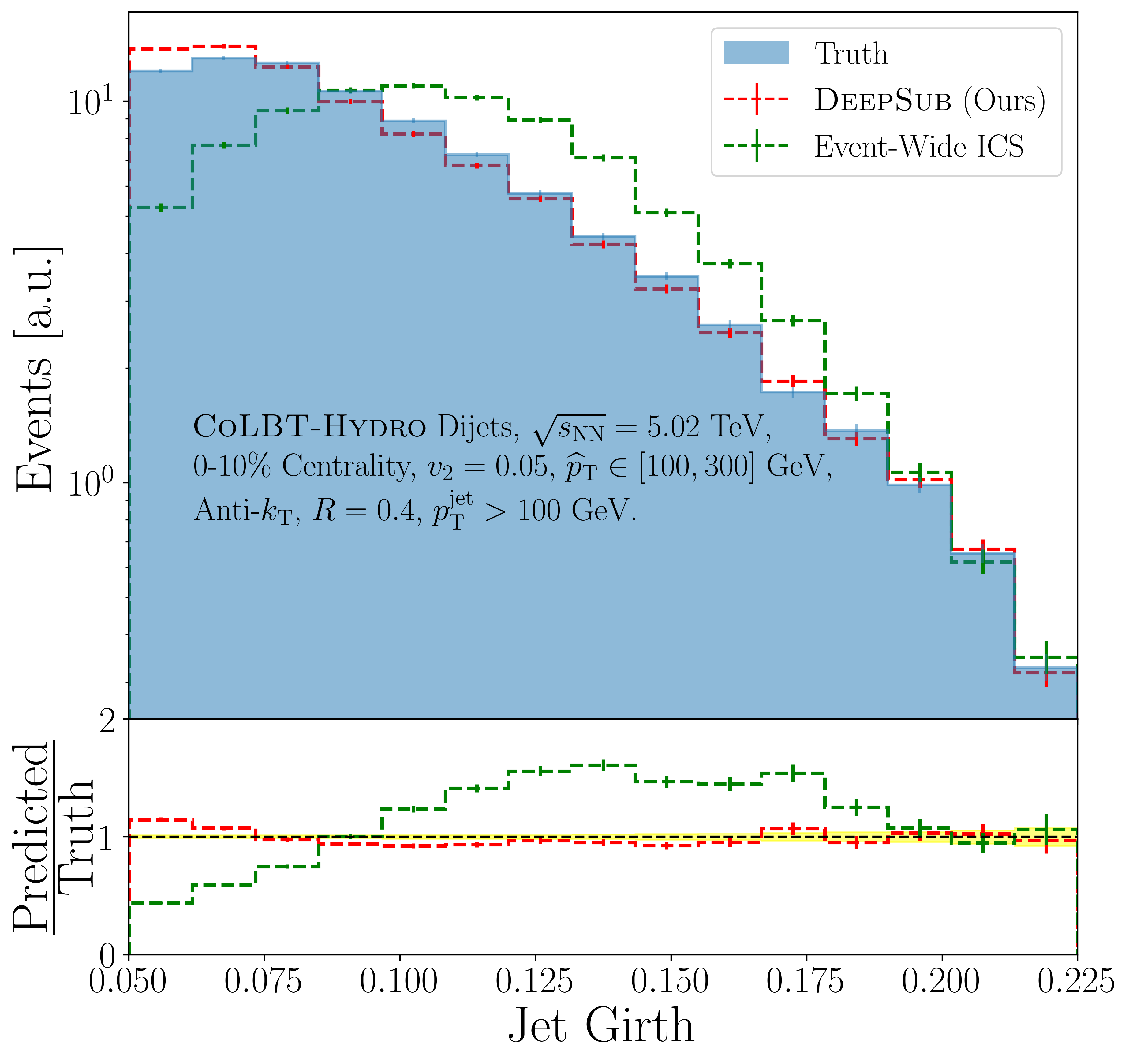}
    \end{minipage}
    \begin{minipage}{0.495\textwidth}
        \includegraphics[width=\linewidth]{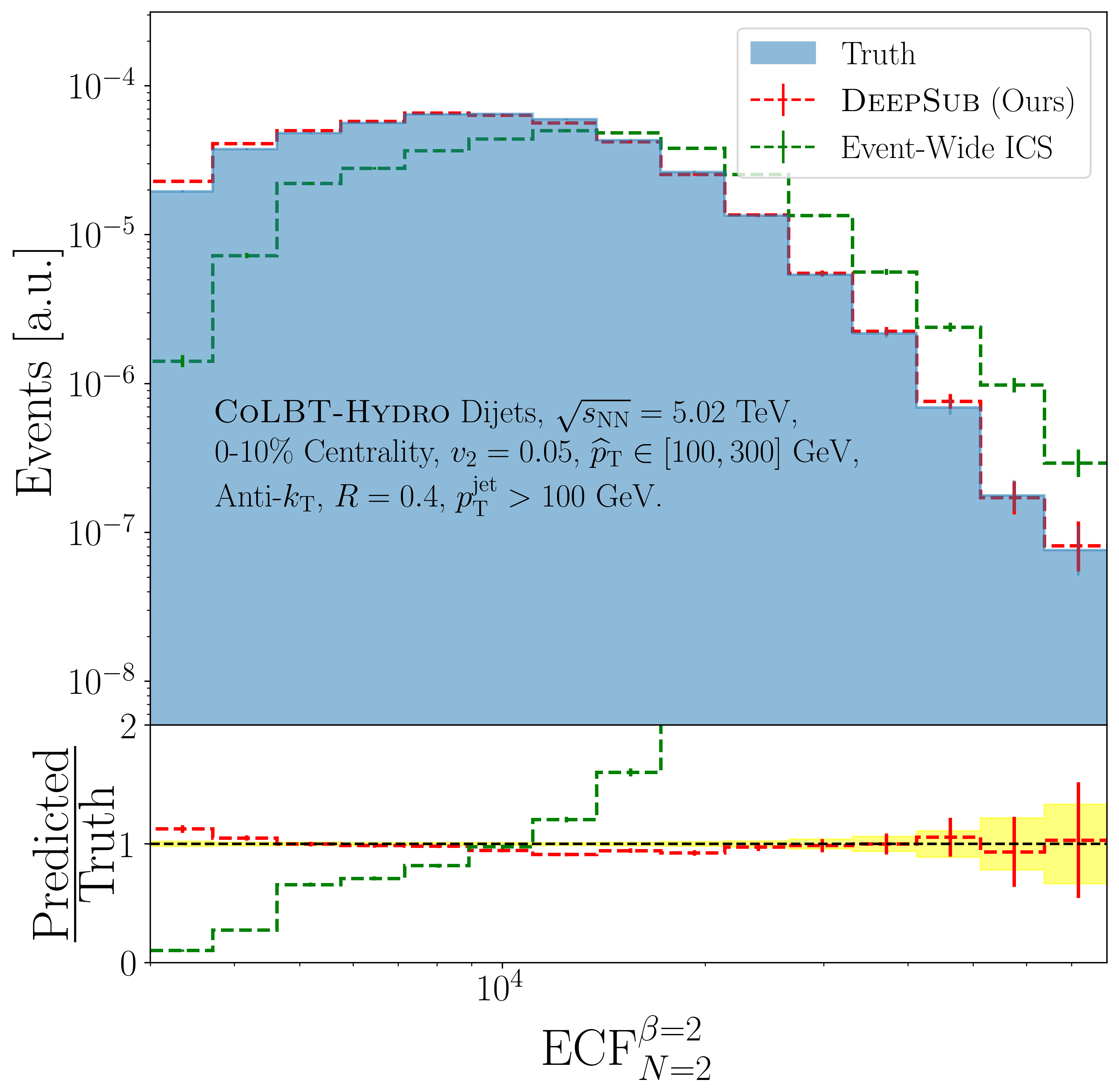}
    \end{minipage}
    \caption{Distributions of jet girth (left), and the energy correlation function (right) for \textsc{DeepSub} (red-dotted), event-wide ICS (green-dashed), and the ground truth (blue-filled) on \textsc{CoLBT-Hydro} dijet events. The error bars represent the statistical uncertainty in each bin. The bottom panels in each plot show the ratio of the reconstructed value of each observable to their truth values and the yellow-shaded region indicates the statistical uncertainty in the truth.}
    \label{fig:rescolbt}
\end{figure*}

The bottom-left panel shows the reconstructed jet girth distribution. At the smallest girth values $\lesssim 0.075$, ICS overshoots the truth by roughly 80-100\%, while \textsc{DeepSub} remains much closer, overpredicting by only 1-5\%. In the mid‐girth region $(0.08 \lesssim g \lesssim 0.12$), ICS transitions from overpredicting to underpredciting, while \textsc{DeepSub} remains within $\approx$ 1-5\% of the truth. Beyond $g \approx 0.12$, ICS increasingly underestimates the yield, falling to $\approx$ 50\% of the truth by $g\approx 0.018$ and down to 10\% by $g \approx 0.25$, whereas \textsc{DeepSub} holds within $\approx$ 5-10\% across the entire high‐girth tail.

Finally, the bottom-right panels show the ECF distribution, which falls by more than six orders of magnitude. At the smallest ECF values $\approx 10^4$, ICS severely underpredicts the yield (ratio $\approx 0.15$) while \textsc{DeepSub} remains within $\approx 5\%$ of the truth. In the midrange (ECF $\in [2\times 10^4, 10^5]$), ICS overshoots by up to 80\%, whereas \textsc{DeepSub} tracks the truth within $\approx$ 1-5\%. At higher ECF ($10^5$ -- $5\times 10^5$), ICS again overestimates by as much as an order of magnitude, while \textsc{DeepSub} continues to stay within $\approx$ 1-5\%. In the extreme tail (ECF $>5\times 10^5$), both methods exhibit statistical fluctuations and \textsc{DeepSub} shows a large dip (to 20\%), indicative of limited counts. Once again, \textsc{DeepSub} reproduces the ECF with significantly higher accuracy than ICS.

While it is encouraging that \textsc{DeepSub} performs so well in Fig.~\ref{fig:resJewel} relative to ICS, it is perhaps not surprising given that it was fine-tuned on the (statistically) same dataset while ICS is generically designed to subtract a variety of underlying event from a signal. Further, it is also conceivable that the model may have over-fitted to model-dependent radiation patterns, which is especially problematic since, in this case, the performance of \textsc{DeepSub} trained on simulation may not translate to real data due to mis-modeling. Thus, as a second set of results, we investigate the performance of \textsc{DeepSub} on event distributions from a generator that it did not encounter during training. For this out-of-distribution validation, we use events from \textsc{CoLBT-Hydro} \cite{Chen:2017zte} simulation with $\widehat{p}_\mathrm{T}\in [100, 300]$ GeV and test them identically to the \textsc{Jewel} events. For brevity, we omit distributions of jet $p_\mathrm{T}$ and mass, and present the more challenging girth and ECF distributions in Fig.~\ref{fig:rescolbt}.

The performance of \textsc{DeepSub} for these distributions continues to be confidence-inspiring, remaining within 1-5\% of the truth, with the exception of tails, despite significant shifts in the spectra themselves. Meanwhile, ICS continues to show large order-of-magnitude deviations from the truth. Although \textsc{DeepSub} was never trained on \textsc{CoLBT-Hydro}, it still outperforms the generic ICS approach by a wide margin, thus demonstrating its robustness to generator shifts, though further systematic studies may be needed to ensure reliable transfer to real data. The main reason one can expect \textsc{DeepSub} to extrapolate to this dataset is that while the models are quite orthogonal, large overlaps exist in radiation patterns and other particle correlations. Therefore, the same latent representations \textsc{DeepSub} learns to distinguish signal from background remain effective. 

Another key advantage of \textsc{DeepSub} is speed. For 80k \textsc{Jewel} events parallelized across a 32-core Intel Xeon (2500 per core), it takes approximately 20 minutes for ICS to run. In comparison, \textsc{DeepSub} running on a fairly modest Nvidia RTX A6000 GPU takes only 3 minutes.
\section{Discussion and Future Work}
\label{disc}
In summary, we have presented a paradigm for constructing a data-driven background subtraction method in heavy-ion environments. This framework is based on a deep learning model that takes as input an event image (noisy image consisting of signal and underlying event) and outputs another image (the reconstructed signal particles). While we have used dijet events at LHC energies as an example, this approach is generalizable to a variety of processes at past, current, and future accelerators. It is also worth reminding that while we chose a handful of observables for their sensitivity to jet substructure, the goal is that whatever observables are used for an analysis, \textsc{DeepSub} should be able to provide an accurate reconstruction.

The long‐term objective of \textsc{DeepSub} is to solve the challenges encountered by experimental collaborations in mitigating the heavy-ion background and to provide a unified analysis framework for both experimentalists and phenomenologists. Non-closure due to background subtraction is often an $\mathcal{O}(1)$ fraction of all systematic uncertainty in a heavy-ion jet substructure analysis. Thus, if we can reduce it to the extent shown in this study, then we could pave the way for precision QCD measurements in heavy-ion collisions. Further, since \textsc{DeepSub}, once trained, is portable, it would be a straightforward one-time exercise for experiments to share their tuned models for background subtraction for anyone to use. 

Although this work represents a major advancement toward this long-term vision, several critical steps remain. First, we must evaluate \textsc{DeepSub}'s performance on other processes such as $\gamma$-jet, $Z$-jet, or minimum-bias events, and also include additional output features such as particle ID information. Next, establishing true universality will require validating the method across a variety of energies, background parameters, and observables. Moreover, for widespread adoption, \textsc{DeepSub} should also provide a flexible, user-friendly interface, on par with that of \textsc{FastJet}, for instance. Lastly, we will need to work with experiments to develop models and facilitate the sharing of these models with the broader community.

With that said, we believe that the progress presented in this Letter serves as proof of concept for \textsc{DeepSub} in establishing a new paradigm for fast, accurate background subtraction in ultra-relativistic heavy-ion environments.

\vspace{-1pt}

\section*{Data and Code Availability}
All data and code, including pretrained models, necessary to reproduce the results presented in this study are publicly available on GitHub at \url{https://github.com/umarsqureshi/DeepSub}. Detailed documentation is provided to facilitate reproducibility and further exploration by the community.

$ $
\vspace{3pt}

\begin{acknowledgments}
The authors express their gratitude to Yi (Luna) Chen (Vanderbilt) for insightful discussions and Michael Kagan (SLAC) for useful feedback on the manuscript. The authors also appreciate the support of the Vanderbilt ACCRE computing facility and the Data Science Institute at Vanderbilt University for providing computing resources, including access to the DGX A100 AI Computing Server. RKE acknowledges funding from the U.S. Department of Energy, Office of Science, Office of Nuclear Physics under grant DE-SC0024660.

\end{acknowledgments}




\bibliography{apssamp}

\end{document}